\documentclass{article}
\usepackage{spconf,amsmath,graphicx,hyperref,float,siunitx}

\title{AiLive Mixer: A deep learning based zero latency automatic music mixer for live music performances}
\name{Devansh Zurale \qquad Iris Lorente \qquad Michael Lester \qquad Alex Mitchell} 
\address{Shure Incorporated}
\ninept
\begin{document}
%\ninept
%
\maketitle
\begin{abstract}
In this work, we present a deep learning-based automatic multitrack music mixing system catered towards live performances. In a live performance, channels are often corrupted with acoustic bleeds of co-located instruments. Moreover, audio-visual synchronization is of critical importance thus putting a tight constraint on the audio latency. In this work we primarily tackle these two challenges of handling bleeds in the input channels to produce the music mix with zero latency. Although there have been several developments in the field of automatic music mixing in recent times, most or all previous works focus on offline production for isolated instrument signals and to the best of our knowledge, this is the first end-to-end deep learning system developed for live music performances. Our proposed system currently predicts mono gains for a multitrack input, but its design along with the precedent set in past works, allows for easy adaptation to future work of predicting other relevant music mixing parameters.

\end{abstract}
\section{Introduction}
\label{sec:intro}

Music mixing is an essential step in the process of producing or performing music. Typically, a mixing engineer is responsible for processing the raw audio tracks of a musical composition, which involves balancing their volume levels and applying various effects such as equalization (EQ), compression, delay, reverb, etc. This task is inherently complex and demands substantial expertise, making it difficult for all musicians to have access to qualified mixing engineers. Therefore, a system to automatically mix music is highly desired.

Since the early work in \cite{dugan1975automatic}, several traditional approaches such as \cite{pachet2000fly, kolasinski2008framework, ward2012multitrack} have been instrumental (pun not intended) towards the progress of the automatic music mixing (AMM) research. Moreover, with recent advancements in artificial intelligence (AI), several deep learning (DL)-based frameworks have been proposed for the AMM task. To the best of our knowledge, most recently proposed systems are targeted towards offline music production. 

In a live music performance, instrumentalists are co-located, causing their instruments to acoustically bleed into each other's microphones. Moreover, it is essential that the music mix is produced with minimal latency to maintain audio-visual synchronization. Although there is some limited work in tackling some of these challenges such as in \cite{perez2009automatic, moffat2019automatic}, to the best of our knowledge no end-to-end DL-based system currently exists for AMM in a live performance setting. In this work we propose such a DL framework for AMM, while focusing on the above mentioned challenges of bleeds and latency.

A potential reason for the lack of previous research in handling bleeds is the lack of data consisting of raw multitrack corrupted with bleeds and the corresponding ground truth mixes. To tackle this, we propose augmenting the data through parametrically simulating bleeds on data consisting of isolated instruments, which is further described in Section \ref{sec:data}.

Architecturally, a recurring theme of many previously proposed DL-based systems such as \cite{ramirez2017deep, martinez2021deep, martinez2022automatic, koszewski2023automatic} is to directly produce the mixed audio from the input multitrack. Doing so limits the users from being able to modify the generated mix, a highly desired functionality, given the subjective and artistic nature of the task. Moreover, many of these works also expect a fixed set and order of instruments as inputs to their models, which further limits the applications of such systems. To tackle these challenges, \cite{steinmetz2021automatic} proposed Differentiable Mixing Console (DMC), which accepts number and order invariant channel inputs to predict mix parameters of typical audio effects for every channel, thus enabling users control over the generated mix. DMC was trained for offline music production and exhibits an inherent latency of $975$ ms. In this paper, we use DMC as the baseline and build upon it to enable handling of bleeds in the input channels and zero latency processing. Although DMC was trained to predict several mix parameters, in this work we only predict mono gain parameters per input audio channel, thus predicting a mono gain mix.

Our primary contributions in this paper are then summarized as follows:
\begin{enumerate} 
    \item We introduce AiLive Mixer (ALM), a modified version of the system flow from DMC, where we propose splitting the processing into two different rates and adding feature conditioning to better support zero latency mix prediction.
    \item We propose neural network architectural modifications such as inclusion of a transformer encoder block to learn inter-channel context and a Gated Recurrent Unit (GRU) block to learn temporal context which are aimed at better handling bleeds in the input and enabling zero latency mix prediction respectively.
    \item We propose data augmentation and training strategies to train the model in presence of bleeds
\end{enumerate}

\section{AiLive Mixer System Design}
\label{sec:system}

\begin{figure*}[ht]
  \centering
  \includegraphics[width=0.8\textwidth]{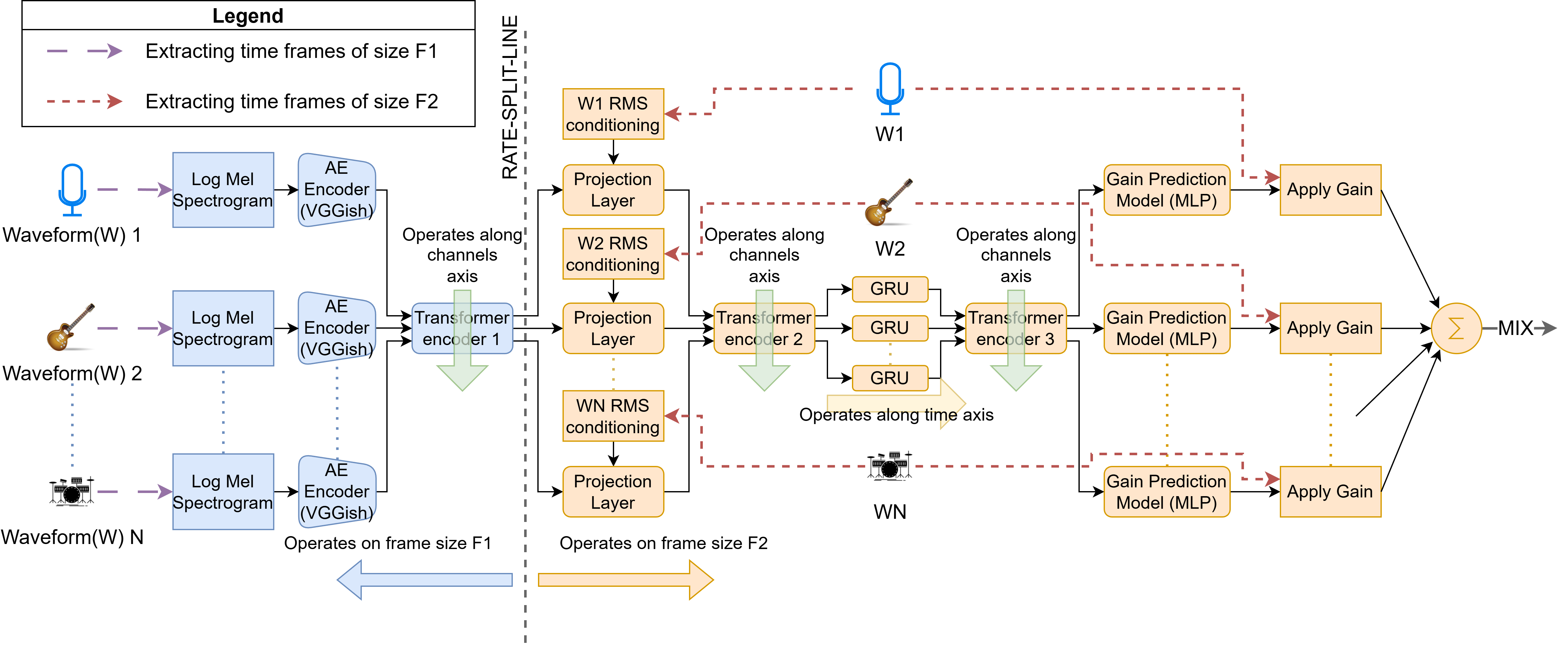}
  \caption{AiLive Mixer - System Overview. Blue blocks to the left of the RATE-SPLIT-LINE operate with a frame size of $F1 = 975$ ms. Orange blocks to the right of the line operate with a frame size of F2. In MR mode, $F2 = 50$ ms, in SR mode $F2 = F1 = 975$ ms.}
  \label{fig:system} 
\end{figure*}

%\begin{figure}[ht]
%  \centering
%  \includegraphics[width=\linewidth]{./Figures/AiLiveMixer-ICASSP}
%  \caption{AiLive Mixer - System Overview. Blue blocks to the left of the RATE-SPLIT-LINE operate with a frame size of $F1 = 975$ ms. Orange blocks to the right of the line operate with a frame size of F2. In this paper we present results when setting $F2 = 50$ ms and when setting $F2 = F1 = 975$ ms}
%  \label{fig:system} 
%\end{figure}

\subsection{Model Architecture}
\label{ssec:architecture}

Fig.\ref{fig:system} shows an overview of our proposed AiLive Mixer (ALM) system. In this every raw audio channel is first passed through an audio embedding model. The extracted features are then passed through several neural network blocks aimed at learning inter-channel and temporal context to predict a mono gain parameter per channel. The predicted gains for all channels are then applied to the respective audio channel waveform before summing together to obtain the resultant mono mix. All the neural network blocks share weights across all channels and operate at one of two frame rates, which is further described in Section \ref{ssec:mrproc}. Below we describe each block in detail.

\vspace{0.5em}
\noindent{\bf Audio Embedding Model:} We use a pre-trained audio embedding model to embed information pertaining to the instrument type, running over a frame size of $F1 = 975$ ms. For this we use the popular VGGish architecture \cite{hershey2017cnn}, but more recent audio embedding models such as \cite{gong2021ast, huang2022masked} could also be used here. To enable VGGish to further embed the extent of bleeds in the respective audio channels, which we hypothesize is critical information required to make accurate gain predictions, we further finetune the VGGish model following the method described in Section \ref{sec:training}.

\vspace{0.5em}
\noindent{\bf RMS Conditioning:} Most audio embedding models, including VGGish, are trained to be indifferent to the amplitude of the input audio signal. The input level of every channel is however valuable information needed to make accurate mix predictions. Thus we compute the Root Mean Square (RMS) of the input audio per $F2$ frame per channel and inject it into the system using a linear projection layer followed by a PReLU activation.

\vspace{0.5em}
\noindent{\bf Temporal Block (GRU):} To introduce temporal context, we use a single layer of the Gated Recurrent Unit (GRU) with a hidden size of $128$ running along the temporal axis. We hypothesize that the importance of temporal context is further highlighted in zero-latency processing where the model is expected to make futuristic predictions for the mix parameters, as further described in Section \ref{ssec:zero-lat}.

\vspace{0.5em}
\noindent{\bf Transformer-encoder Blocks:} The mix parameters for a given audio channel, not only depend on the channel itself, but are also expected to depend on all the other channels present in the given multitrack. We use transformer-encoder blocks, which run along the channels axis, to learn inter-channel context using its multiheaded self-attention mechanism. We use only a single layer of the transformer-encoder and use $2$ heads per block. We place these blocks at three different points within the system –- 1. right after the audio embedding model, to learn context pertaining to the instrument types, 2. right after RMS conditioning, to learn context relating to the relative levels of the channels and 3. right after the temporal GRU block to learn the inter-channel context of the temporally informed embeddings. Through experimentation, we found that placing such dedicated transformers worked better than placing a single transformer-encoder with more layers at a single point in the system.

\vspace{0.5em}
\noindent{\bf Gain Prediction MLP:} Finally we use a Multi-Layer Perceptron (MLP) to predict the gain parameter for every channel. The MLP consists of $3$ hidden layers having sizes of $128$, $64$ and $32$ and a uni-dimensional output layer corresponding to the predicted gain. Every layer uses a PReLU activation function, except the final layer which uses a ReLU activation.

\subsection{Multi-Rate (MR) Processing}
\label{ssec:mrproc}

\begin{figure}[t]
  \centering
  \includegraphics[width=0.95\linewidth]{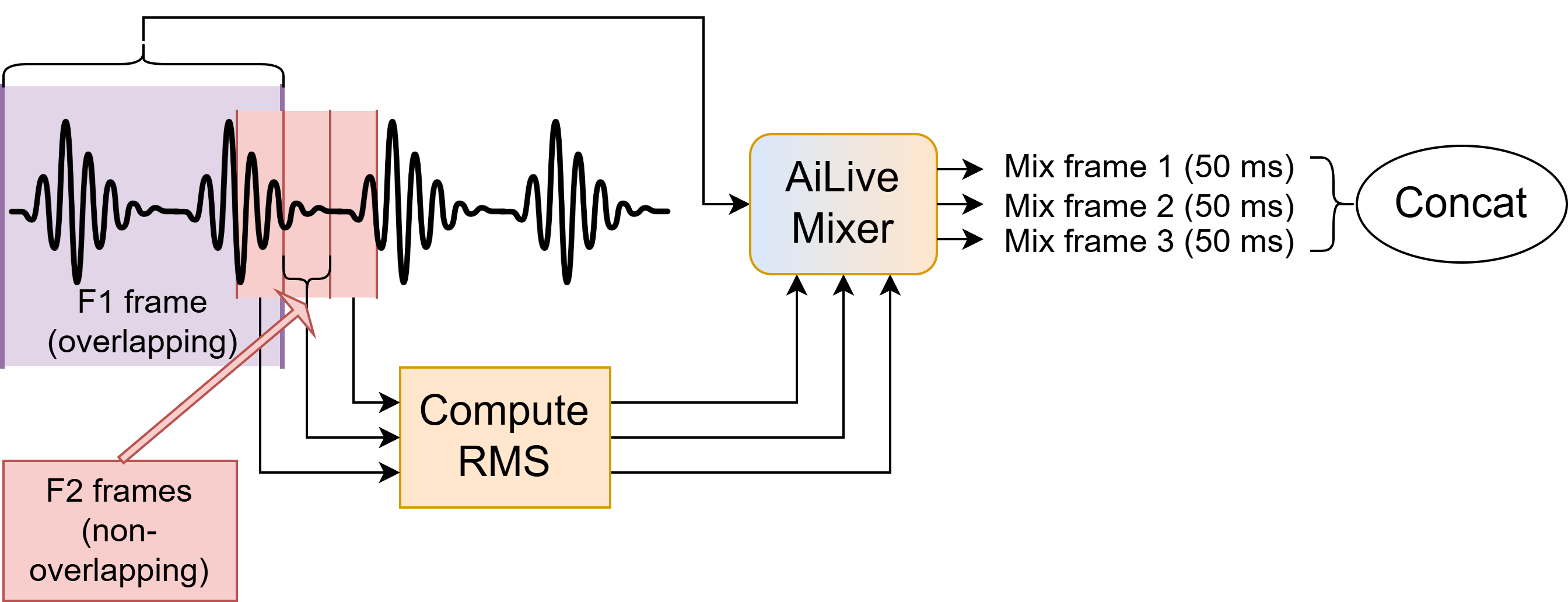}
  \caption{Multi-Rate Processing for a single channel.}
  \label{fig:framing} 
\end{figure}

To enable low latency processing, we split the system into two frame rates. Typically audio embedding models require large temporal context to embed relevant information. Thus we extract longer frames to pass through the audio embedding model, in our case $975$ ms as expected by VGGish. Transformer-encoder 1 also operates on this frame size. Features such as RMS on the other hand, do not need such long receptive fields. Thus, we provide the capability to run the rest of the processing on shorter frame sizes, $50$ ms in our case. By splitting the processing into two rates, the system can make gain predictions faster, without having to wait for the longer audio frames to become available. The frame processing flow for a single channel is shown in Fig.\ref{fig:framing}. In this, we first extract an F1 frame ($975$ ms) and several non-overlapping F2 frames ($50$ ms). ALM then predicts gains per F2 frame, thus reducing the latency of the system to the F2 frame size. Theoretically a new F1 frame can be extracted for every new F2 frame, which would mean striding the F1 frame by the size of F2 frame. However, to manage computational load, we use the same F1 frame for several F2 frames, before extracting a new F1 frame. For the specific case of F2 frames being $50$ ms we stride the F1 frame by $6$ F2 frames, which accounts to a F1 frame stride duration of $300$ ms.

Note that setting F2 frame size to be equal to the F1 frame size would mean that there are no rate splits. In this work we also provide results for this case in Section \ref{sec:results}, which is referred to as Single-Rate (SR) processing. In the SR case, both F1 and F2 frames are $975$ ms and non-overlapping.

\subsection{Zero Latency Processing}
\label{ssec:zero-lat}

The proposed multi-rate (MR) processing although reduces the latency of the system from $975$ ms to $50$ ms (or lower for a lower F2 frame size), it doesn't directly translate to zero latency. To then achieve true zero latency, we train our model to predict gain values for the upcoming audio. Specifically, when an F2 frame is extracted and processed, the model generates a gain prediction that is then applied to the subsequent F2 time frame (one frame ahead), to obtain the predicted mix. This approach ensures that during inference, as new input audio arrives, the model has already predicted the desired gain value for it, allowing immediate application and thereby enabling the system to operate with genuine zero latency. Although this can theoretically be achieved without MR processing, we hypothesize that through MR processing the model has to only predict $50$ ms into the future, as opposed to having to predict $975$ ms into the future, as would be the case with SR processing. This training strategy of using zero latency processing will be referred to as zero latency training.

\subsection{How ALM differs from DMC}
\label{ssec:alm-vs-dmc}

DMC consists of the VGGish audio embedding model followed by concatenating the embeddings per channel with the mean of audio embeddings across channels to inject inter-channel context. These are then passed through an MLP model to predict the mix parameters. The RMS conditioning to embed input level information, the transformer-encoders for learning the inter-channel context, the GRU for learning temporal context along with the multi-rate and zero-latency processing are what puts ALM apart from DMC. Additionally, while DMC was proposed for predicting mix parameters for offline production, ALM is trained to handle bleeds (further described in Section \ref{sec:data}). DMC however was also trained originally to predict mix parameters that go beyond gains, such as parameters for panning, EQ, compression, delay and reverb. In this work, we trained ALM to only predict mono gains. However, given the precedence set by DMC, ALM could easily be extended to also predict such other mix parameters, which is left for future work.

\section{Training Data \& Data Augmentation}
\label{sec:data}

In this work we use MedleyDB\cite{bittner2014medleydb,bittner2016medleydb} to train our model, which is a dataset that consists of raw multitrack recordings and the corresponding human-made mixes. When it comes to live performances, several factors govern the amount and kind of bleeds that every track receives, such as the dimensions of the performance space, reverb levels of the performance space, relative positions of the performers, relative acoustic levels of the instruments, distance of instruments from the microphones. Collecting a dataset that encompasses sufficient combinations of all the above factors is impractical. Thus, in this work we use isolated instrument tracks from MedleyDB and artificially simulate bleeds into these raw tracks. To this end we built a parametric tool for bleed simulation using pyroomacoustics\cite{scheibler2018pyroomacoustics}, a python library to simulate room responses. We randomize the bleed parameters on the fly during training, thus generating a large range of bleed scenarios during training. Additionally, we also randomize the input levels for all tracks going into the system on the fly during training. This is done both after and before bleed simulation to support a wide range of input levels as well as relative instrument level balances respectively.

Given the impracticality of getting ground truth mixes for all bleed-simulated versions, for all such variants of a given multitrack, we use as ground truth the same mix that's available in MedleyDB, which was made for the isolated tracks. We normalized the ground truth mixes for all songs to $-6$ dB FS peak.

\section{Training Methods}
\label{sec:training}

We used songs with $<=8$ raw tracks from MedleyDB, those which consisted of isolated instruments. We used a $80/20$ train/val split, thus corresponding to a total of $43$ songs with $35$ songs for training and $8$ songs for validation. Note that although during training we used songs with isolated raw tracks with simulated bleeds, at inference we use real live performances which consisted of raw tracks with naturally occurring bleeds, results for which are presented in Section \ref{sec:results}. We trained the model for $5000$ epochs using an AdamW optimizer, where one epoch corresponds to a full pass through all the songs in the training set while randomly sampling $20$ seconds snippets at every epoch. We started with an initial learning rate of $0.001$ and used a multi-step learning rate scheduling, reducing the rate by a factor of $10$ at epochs $100$, $1000$, $2500$. 

For the loss function, we use the multi-resolution STFT loss\cite{yamamoto2020parallel}, using window sizes of $440$, $884$ and $3528$ which correspond to about $0.01$, $0.02$ and $0.08$ seconds respectively, $25\%$ hop size for all windows and fft sizes of $512$, $1024$ and $4196$ for each window respectively. We used the auraloss library\cite{steinmetz2020auraloss} to compute this loss.

For VGGish finetuning, we first freeze the weights of VGGish for the first $100$ epochs and then unfreeze its weights, keeping the learning rates of VGGish and the rest of the model to be the same. We found that this approach of finetuning provided the best results.

\section{Experiments}
\label{sec:experiments}

To demonstrate our contributions we trained $4$ models:
\begin{enumerate}
\label{list:models}
	\item \textbf{ALM-MR (ours):} Multi-Rate Processing using ALM architecture trained using bleed simulations and zero latency
	\item \textbf{ALM-SR (ours):} Single-Rate Processing using ALM architecture trained using bleed simulations and zero latency
	\item \textbf{DMC-B-0L (hybrid):} DMC model architecture, but trained using bleed simulations and zero latency
	\item \textbf{DMC-OG (baseline):} DMC as it was proposed -- using DMC architecture, trained without bleed simulations or zero latency
\end{enumerate}

Note: To provide a fair comparison with ALM models, we trained both the DMC variants to only predict mono gains.

We then evaluated all models on multitrack recordings of live performances, and generating the mixes with zero latency. Additionally, the raw tracks that were input to these models during inference were intentionally poorly gain staged to demonstrate a novice user.

Given the subjective nature of the task, objective analysis of mix quality is a challenging subject of research and no real objective metric currently exists that is proportional to the perceptual quality of the mixes\cite{steinmetz2020learning}. Thus we conducted a subjective study to perceptually evaluate the models. The study consisted of $15$ people with critical listening skills and all working in the field of audio, including some musicians and some with a background in music mixing. Given that the application of this model would be in live performances, we didn't restrict the participants to only audio engineers. We used the Web Audio Evaluation Tool\cite{jillings2016web} using the APE test design\cite{de2014ape}. In this, we picked songs from an internal dataset of live performances and we played  $20-30$ s sections of every song. $8$ such segments were used for this study. The songs were chosen to encompass different genres of music and recorded in different scenarios. For each song, participants were presented with mixes generated using all the $4$ model variants along with a fifth mix -- the raw mix (denoted as RAW) which was the sum of all input raw tracks. Note that to mimic a live mixing scenario no normalization was applied to the generated results. 

\section{Results \& Discussion}
\label{sec:results}

\begin{figure}[t]
  \centering
  \includegraphics[width=0.8\linewidth]{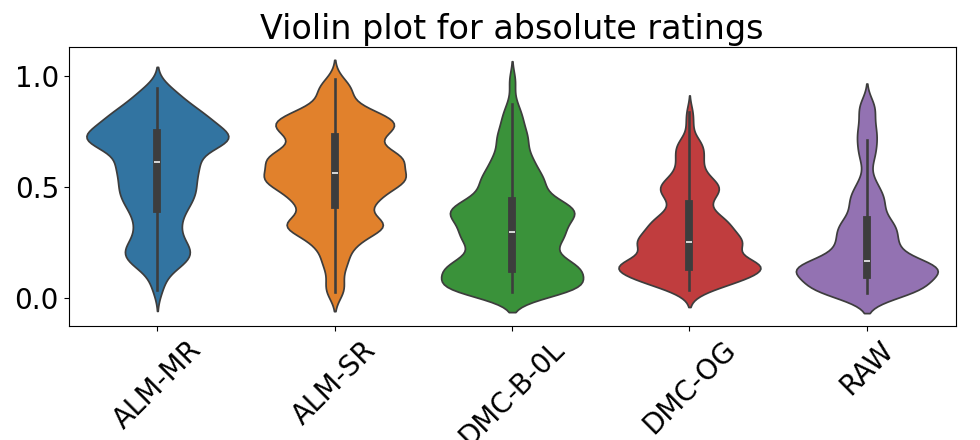}
  \caption{Violin $+$ Box: Absolute Ratings}
  \label{fig:subj-eval} 
\end{figure}

\begin{figure}[t]
  \centering
  \includegraphics[width=0.8\linewidth]{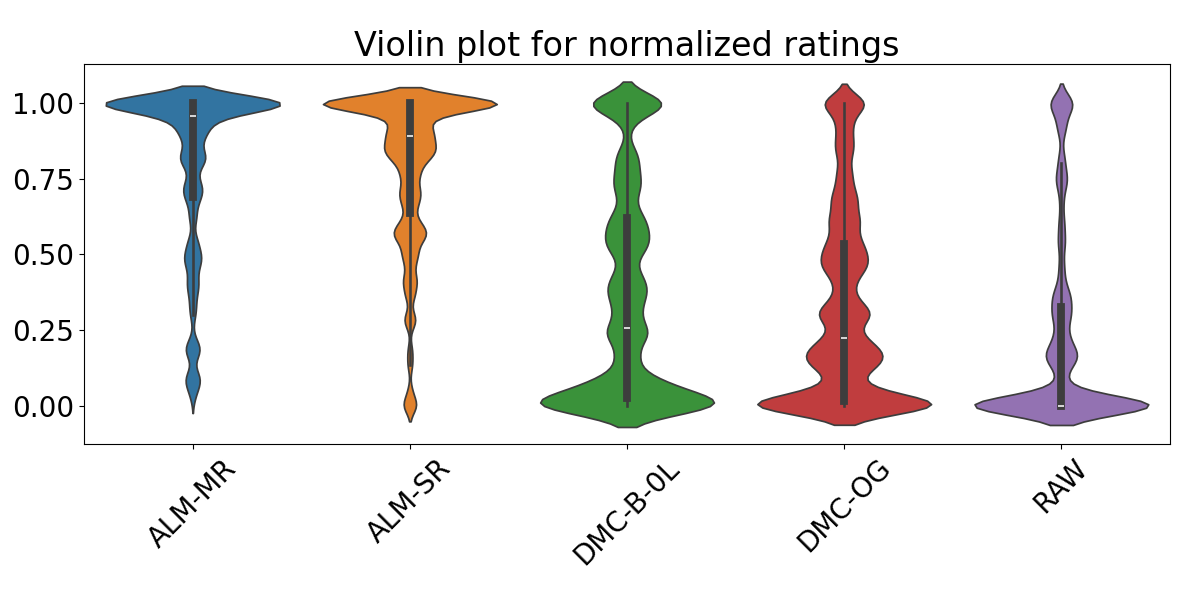}
  \caption{Violin $+$ Box: Ratings normalized per song per person}
  \label{fig:subj-eval2} 
\end{figure}

To summarize our findings from the listening test, we provide violin plots for the model ratings in Figure \ref{fig:subj-eval}. Please also note the box plots that are embedded within the violin plots. The plots suggest that overall, both the ALM models outperformed the DMC models as well as the raw mix. The ratings for ALM-MR are clustered towards a higher rating of $\sim 0.75$ while the ratings for DMC-OG, DMC-B-0L and RAW are all clustered towards lower scores of $< 0.5$. As for ALM-SR, the ratings are clustered higher than the DMC models and RAW, however the ratings are more spread out as compared to ALM-MR's more focused clustering in the higher ratings region. Although the difference between ALM-MR and ALM-SR might be perceived as subtle, we hypothesize that ALM-MR could provide significant improvements for the future work of predicting mix parameters other than mono gain which our model currently predicts. Additionally, to demonstrate a relative rank, we also provide Figure \ref{fig:subj-eval2}, where the violin plots were computed for the ratings after min-max normalizing the ratings per song and per participant. Looking at the plots, the ranking could be summarized as ALM-MR $>$ ALM-SR $>$ DMC-B-0L $>$ DMC-OG $>$ RAW, which is in line with our hypothesis.

To formalize these results further, we also performed a Kruskal Wallis H test\cite{kruskal1952use} on the absolute ratings, which resulted in $H = 156.485, p = \num{8.293e-33}$, showing that there is a statistically significant difference between some of the models. We then performed the Conover's test to obtain pairwise p-values for all model combinations. In summary, the p-values suggested that both ALM models are statistically significantly different from the rest of the models with p values in the order of $\num{1e-15}$ - $\num{1e-24}$. DMC-B-0L also showed statistical difference vs RAW. There wasn't enough evidence to suggest significant difference between ALM-MR vs ALM-SR or DMC-OG vs DMC-B-0L or DMC-OG vs RAW.

We also provide audio results for the models at \url{https://dzurale.github.io/ailive_mixer_icassp2026/}. The snippets presented in these results were taken from MedleyDB and consists of songs that contain naturally occurring bleeds in their raw tracks. Note that these mixes were all generated with $0$ latency.

\subsection{Audio Results Discussion}
\label{ssec:audio-results}

In general, we found the DMC models to consistently exhibit abrupt gain jumps between successive audio frames, which can be attributed to the missing temporal network in their model architecture. Even when trained with bleeds and with zero latency, the lack of the temporal model hampers their ability to make consistent gain predictions. ALM-SR also tends to exhibit some gain jumps, but far less often than the DMC models, particularly at sudden introductions of instruments. This can be attributed to the slow reaction time of ALM-SR due to its large frame size. ALM-MR almost never results in any abrupt gain jumps which is also likely why it is most consistently highly rated in the listening study. This also likely explains the difference between the violin plots of ALM-MR vs ALM-SR where ALM-MR is mostly concentrated around higher ratings, while ALM-SR is more spread out. ALM-MR's consistency was also noted by some participants through comments that we also collected as part of the listening study. Another finding that we've heard in the model outputs is that the ALM-MR model most consistently results in mixes that are free from clipping, while the other models occasionally result in mixes that go beyond the clipping range. This is critical for systems to be used in live performances as normalization may not be possible. Lastly, we have also seen ALM-MR to result in better mix predictions for percussive sounds such as a bass guitar as compared to the SR models, including ALM-SR as well as the DMC models. This is likely because the ALM-MR model adapts to RMS computed over shorter frames, allowing for transients in the input signals to be better modeled.

\subsection{Conclusion}
\label{ssec:conclusion}
In this paper, we introduced a system for automatic music mixing tailored for live performances, effectively addressing track bleed issues and achieving zero latency mix prediction. To this end, we implemented a multi-rate processing strategy (see Section \ref{ssec:mrproc}) and a zero-latency processing approach (see Section \ref{ssec:zero-lat}) and we developed a bleed simulation strategy for model training (detailed in Section \ref{sec:data}). Critical architectural enhancements to the baseline DMC model were also incorporated to facilitate these advancements, as discussed in Section \ref{ssec:architecture}. Our subjective listening study and accompanying audio results affirm that our proposed model outperforms the baseline DMC model. Although currently focused on mono gain predictions, we anticipate that extending our ALM-MR model to predict additional mix parameters will further amplify its effectiveness in future developments.

\section{Acknowledgements}
\label{sec:ack}
This work is based on concepts from a filed provisional patent.

\bibliographystyle{IEEEbib}
\bibliography{strings,refs}

\end{document}